\documentclass[a4paper,man,natbib,floatsintext]{apa6}
 
\usepackage[english]{babel}
\usepackage[utf8x]{inputenc}
\usepackage{amsmath}
\usepackage{graphicx}
\usepackage{algorithm}
\usepackage[noend]{algpseudocode}

\title{Curiosity-Driven Recommendation Strategy for Adaptive Learning via Deep Reinforcement Learning}
\shorttitle{Curiosity-Driven Recommendation Strategy}
\author{Ruijian Han, Kani Chen and Chunxi Tan*}
\affiliation{Department of Mathematics,
	Hong Kong University of Science and Technology,
	Clear Water Bay, Kowloon, Hong Kong }

\note{*Corresponding Author: ctanac@connect.ust.hk (Chunxi Tan)}
\authornote{
	
	Correspondence concerning this article should be addressed to Chunxi Tan, Department of Mathematics, Hong Kong University of Science and Technology, Clear Water Bay, Kowloon, Hong Kong.  E-mail: ctanac@connect.ust.hk}

\abstract{

    The design of recommendations strategies in the adaptive learning system focuses on utilizing currently available information to provide individual-specific learning instructions for learners. As a critical motivate for human behaviors, curiosity is essentially the drive to explore knowledge and seek information. In a psychologically inspired view, we aim to incorporate the element of curiosity for guiding learners to study spontaneously. In this paper, a curiosity-driven recommendation policy is proposed under the reinforcement learning framework, allowing for a both efficient and enjoyable personalized learning mode. Given intrinsic rewards from a well-designed predictive model, we apply the actor-critic method to approximate the policy directly through neural networks. Numeric analyses with a large continuous knowledge state space and concrete learning scenarios are used to further demonstrate the power of the proposed method.
	  \\}
  

\begin{document}
\maketitle

\section{1. Introduction}

 
Adaptive learning is an educational method implemented through computerized algorithms, which orchestrates personalized learning instructions to meet the unique learning needs of individuals. 
Because of its achievability at scale with the advanced Internet access, adaptive learning has brought the new wave in the E-learning field \citep{sleeman1982intelligent,Wenger1987}. The crux of a good adaptive learning system is to design a recommendation system that sequentially outputs customized recommendations, finally leading to the best suitable learning path for every single learner. 

Such a decision-making problem involves a dynamic and interactive environment \citep{chen2018recommendation}. Specifically, with full utilization of learning experience data, psychometric assessment models keep track of the learner's proficiency levels on knowledge points, i.e., knowledge states and then a good recommendation strategy selects the most appropriate action to maximize overall gain according to the current knowledge statue. In the perspective of learners, they take actions each time, receive rewards and transit to the next knowledge state. For a practical learning system with existences of assessment errors, unknown transition kernel and complex rewards, there is a small but growing literature dedicated to the design of recommendation strategies. \cite{chen2018recommendation} proposed the Q-learning algorithm that approximates the objective function by a linear model. Then \cite{tan2019recommendation} raised a model-free solution based on deep Q-network for a sufficiently complex environment model. Beyond the efficiency required for recommendation strategies, we actually wonder if the hand-designed reward settings satisfy the learning needs of learners in a psychologically inspired view and if the proposed recommendation strategies provide learners with a both rewarding and enjoyable learning experience. 

Reinforcement Learning (RL), one of the central topics in artificial intelligence \citep{kaelbling1996reinforcement}, builds a bridge to connect the psychological need of learning behaviors with efficient recommendation strategies in the adaptive learning system. Reinforcement learning refers to goal-oriented techniques, where a simulated agent interacts with the environment by taking actions with a goal of maximizing total rewards and the environment often follows the setup as a Markov decision process(MDP). From achieving an expert human level in Go \citep{silver2016mastering} to defeating amateur human teams at Dota 2, deep reinforcement learning techniques reach unprecedented success in challenging domains. 

The current work of this paper focuses on developing a curiosity-driven recommendation strategy for the personalized learning system based on deep reinforcement learning algorithms. Recognized as a critical impetus behind human behaviors \citep{loewenstein1994psychology}, curiosity is a desire of our nature towards complete knowledge and information-seeking in terms of learning. 
Our objective is to improve the recommendation strategy by incorporating curiosity to further plan an both intrinsically motivated and efficient learning path for the individual.
Therefore, a predictive model is constructed and knowledge points with high prediction accuracy tend to be the ones that learners are familiar with and may feel less curious about. Motivated by \cite{pathak2017curiosity}, a feed-forward neural network is built to generate prediction errors of knowledge points, which further serve as intrinsic rewards to encourage learners to follow their inner curiosities and explore. With curiosity rewards, we approximate the optimal recommendation policy based on the actor-critic framework \citep{barto1983neuronlike,konda2000actor,mnih2016asynchronous}. The proposed method can optimize the objective function using far less computational resources and scale up comparatively easily in handling big data.



The rest of the paper is organized as follows. In Section 2, we formulate the problem under the RL framework. Then, a curiosity-driven recommendation strategy specially designed for the personalized learning system is proposed in Section 3. In Section 4, concrete simulated experiments are provided to examine the proposed policy in the more realistic learning scenarios, followed by the discussion in Section 5.

\section{2. Problem Formulation}

In this section, we address challenges by proposing an integrated online learning procedure and elaborate on the problem of optimizing the learning action execution order in a systematic way.

Consider a learner having $K$ knowledge points to master in a course within finite time step $t = 0,1,2\dots,T$. The knowledge point is defined as a piece of knowledge that can be explicitly defined and widely accepted. Let $\boldsymbol{s}=(s_{1}, s_{2},\dots, s_{K})$ be a $K$-dimensional latent knowledge state of a learner, where $s_{i}(t)$ is the corresponding mastery attribute on the $i$th knowledge point at $t$. 
As one of indispensable modules in the adaptive learning system, the psychological assessment model will measure knowledge state once the learner takes a learning action. It utilizes the item responses of students in the test and delivers the assessment result $\boldsymbol{\hat{s}}$ as the basis in further recommendation. 
Specifically, these assessment models including the multidimensional three-parameter logistic IRT model (M3PL; \cite{reckase2009multidimensional}), noisy AND gate model (DINA; \cite{junker2001cognitive}) and other cognitive diagnosis models, are widely adopted for different test settings. Although not studied in this paper, the parameters of assessment models in the simulations are assumed to be well-calibrated by historical data.

\begin{figure}[htb]
	\centering	
	\includegraphics[width=.86\linewidth]{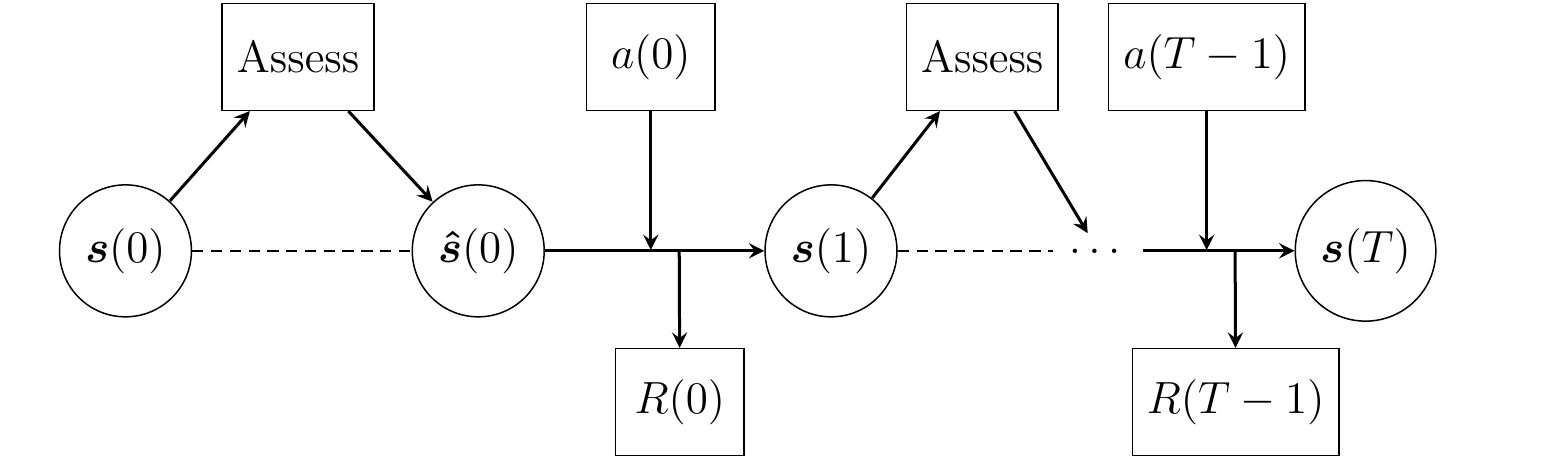}
	\caption{The flow chart of an adpative learning system: $(1)$ the knowledge state $\boldsymbol{s}(t)$ is partially observed by item responses, estimated as $\boldsymbol{\hat{s}}(t)$ in the assessment model; $(2)$ given $\boldsymbol{\hat{s}}(t)$, action $a(t)$ is recommended following the policy $\pi$; and $(3)$ according to the learning action $a(t)$, the unknown transition model determines the next knowledge state $\boldsymbol{s}(t+1)$.}
	\label{flow}
\end{figure}

With the aid of the assessment model, the learning flow path in the adaptive learning system can be represented in Figure \ref{flow}. As it shows, the learner transits into the next state each step by taking certain learning actions:
$$	\boldsymbol{s}(t)\xrightarrow{a(t)} \boldsymbol{s}(t+1).$$
Such a state transition model is assumed to be known in \cite{chen2018recommendation}.
As a matter of fact, the state transition model is not known a \textit{priori} but evolves with the recommendation strategy. The problem is actually formulated in such a way that the transition probability changes along with the action changes. From this perspective, the unknown transition model completely characterizes the most important aspect of the dynamics in the environment, which associates the improvement on the corresponding mastery attributes with certain learning actions. 

Therefore, the desired recommendation policy, denoted as $\pi$, is required to maximize total rewards and thoroughly capture the specific unknown transition model for every single learner in the meantime. Being a mapping function from the state space to a probability distribution over actions, $\pi$ assigns a probability of taking action $a(t)$ given state $\boldsymbol{s}(t)$, i.e., $\pi(a(t)|\boldsymbol{s}(t)) = P(a=a(t)|\boldsymbol{s}=\boldsymbol{s}(t))$. For a MDP, it can be shown that any decision made at $t$ can be based solely on $\boldsymbol{s}(t)$ rather than $\{\boldsymbol{s}(0),\boldsymbol{s}(1),\dots,\boldsymbol{s}(t-1)\}$. Similar learning procedures can be found in \cite{chen2018recommendation,li2018optimal,doi:10.1111/bmsp.12144,tan2019recommendation}.

Following $\pi$, the learner takes a recommended action $a(t)$, transits into next knowledge state $\boldsymbol{s}(t+1)$ and receives a scalar reward $R(t)$ as an immediate feedback. This process continues until the agent observes the terminal state $\boldsymbol{s}(T)$ in a finite time horizon up to $T$. Such interactions with the unknown environment can help agent learn to alter its own decision in response to rewards received.
Every rollout of a learning trajectory can accumulate rewards from the environment, resulting in the overall gain $\sum_{t=0}^{T-1}{R(t)}$. For the ideal environment with the known transition model and perfect assessments for knowledge states, the best sequence of actions that maximize total rewards during the learning process can be determined by the optimal policy $\pi^*$, that is $\pi^* = \arg\max_{\pi}E^{\pi}(\sum_{t=0}^{T-1}{R(t)})$. The expectation accounts for the uncertainties with respect to the environment and the stochastic policy during the whole learning trajectory. 

Our goal is to identify the best treatment of the learning action space in conjunction with the curiosity-driven exploration of the environment. In the next section, we illustrate how we combine curiosity into the reward setting and then approximate the optimal policy in a practical environment. 



\section{3. Curiosity-Driven Recommendation Strategy}

In a psychological inspired view, we combine the element of curiosity as a driving force of exploration to improve an efficient recommendation strategy adaptive to varied learning paces and intrinsically motivating learners simultaneously. In this section, we propose the algorithm that consists of two subsystems: (1) a predictive model is built to provide its prediction error as the curiosity reward signal \citep{schmidhuber1991curious, chentanez2005intrinsically, pathak2017curiosity}; (2) given the curiosity rewards, we directly optimize the policy $\pi$ based on the actor-critic framework \citep{barto1983neuronlike, konda2000actor,mnih2016asynchronous}.

\subsection{3.1. Predictive Model: Curiosity Reward} 
The immediate reward stipulates the way we wish the learner to accomplish. Different learning goals determines different reward settings. We have seen various pre-specified rewards in previous works, which we refer to as extrinsic rewards. For example, \cite{chen2018recommendation} and \cite{doi:10.1111/bmsp.12144} adopted 
$R(t) = \sum_{k=1}^K w_k(s_k(t+1)-s_k(t))$ that pushes the learner to master as many knowledge points as possible, where $w_k$ is the weight for the $k$-th knowledge point. \cite{li2018optimal} and \cite{tan2019recommendation} added the current learning time $t$ in the reward setting to pursue the learning efficiency. With the learning experience data that contains the sequence of observed states, actions and rewards, extracting a reward intrinsic to the agent seems to be more reasonable to reflect the inner curiosity of the learner.

Given current estimated knowledge state $\boldsymbol{\hat{s}}(t)$ and learning action $a(t)$, we consider to predict the next knowledge state by a fully-connected feed-forward neural network $f(\cdot)$ with a set of parameters $\boldsymbol{\theta_p}$, i.e., 
$$\boldsymbol{\tilde{s}}(t+1) = f(\boldsymbol{\hat{s}}(t),a(t);\boldsymbol{\theta_p}),$$
which outputs the next knowledge state predictor $\boldsymbol{\tilde{s}}(t+1)$. The neural network arranges layers and units in a chain structure. Thanks to its flexible architecture, a feed-forward network with one layer can even represent functions of increasing complexity \citep{cybenko1989approximation}. Readers are referred to \cite{goodfellow2016deep} for a comprehensive review about neural networks.

We then apply the mini-batch iteration to update $\boldsymbol{\theta_p}$. Consider the learning experience data at each time step are stored as $e_t=\{\boldsymbol{\hat{s}}(t), a(t),\boldsymbol{\hat{s}}(t+1)\}$ in a memory pool $D$ with size $N$, $D=\{e_1,\cdots,e_N\}$. In each iteration, a batch of samples of size $n$ is drawn at random from $D$. With independent samples $e_{j, j\in|n|}$, the parameter $\boldsymbol{\theta_p}$ is optimized by reducing errors between the next knowledge state estimate $\boldsymbol{\hat{s}}(j+1)$ and the predictor $\boldsymbol{\tilde{s}}(j+1)$, that is to minimize
$$L_p(\boldsymbol{\theta_p})=\sum_{j=1}^{n}\lVert\boldsymbol{\hat{s}}(j+1)-\boldsymbol{\tilde{s}}(j+1)\lVert_2^2.$$
With the constant update of the data in the memory pool, the predictive model follows the learner's knowledge statue and keeps being calibrated at each time step. For time step $t$, the immediate reward $R(t)$ is calculated by 
\begin{equation}
\label{reward}
R(t) = \lVert \boldsymbol{\hat{s}}(t+1)-\boldsymbol{\tilde{s}}(t+1)\lVert_2^2.
\end{equation}

\begin{figure}[htb]
	\centering	
	\includegraphics[width=.66\linewidth]{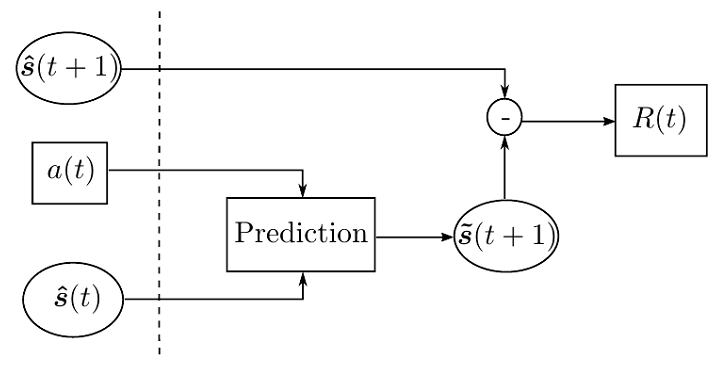}
	\caption{Given $\{\boldsymbol{\hat{s}}(t),a(t),\boldsymbol{\hat{s}}(t+1)\}$, the predictive model outputs the curiosity reward $R(t)$.   }
	\label{prediction}
\end{figure}

We show the flow chart of the predictive model in Figure \ref{prediction}. 
The predictive model is essentially to learn from learning experience data with the aim towards estimating the learner's familiarity with all knowledge points of the moment. High prediction error delivers a good signal to the agent and encourages the learner to take the corresponding action, resulting in reducing the uncertainty in predictions on the consequence of the learner's behaviors. For example, when a learner touches a new area of knowledge space, a high reward will be generated and further serves as a bonus for this exploration in the learning trajectory. In other words, the curiosity that is partially reflected in the reward will push the algorithm to favor transitions with high prediction error and hence better explore the knowledge space. In this process, timely updates of the predictive model guarantee a firm grasp of one's familiarity with knowledge points. Therefore, a more accurate predictive model delivers an opportune reward and prevents the learning from getting stuck.


\subsection{3.2. Policy Learning Based on the Actor-Critic Framework}
As the learner's exploration for knowledge goes deeper, the predictive model generates dynamic curiosity rewards that may tend to be smaller over time. Due to dynamic rewards in relatively large state space, function approximation methods including approximate dynamic programming \citep{thrun1993issues, powell2007approximate} and deep Q network \citep{mnih2013playing,mnih2015human} may be problematic. Therefore we leverage the actor-critic method as our algorithmic backbone to approximate the policy via the neural network, which is comprised of two closely related processes. One is the actor actuator, estimating the current policy; another one is the critic evaluator, aiming at evaluating the quality of the policy. Consulting $\pi$ from the actor, the action $a(t)$ is executed and the next knowledge state along with the feedback from the environment will be delivered to the critic. 

\subsubsection{Actor-Critic}
We elaborate on the actor-critic method below. For the actor, we aim to approximate the policy $\pi$ via a parameterized non-linear function $\pi(a|\boldsymbol{\hat{s}};\boldsymbol{\theta_{\pi}})$. As a universal approach for representing functions, a feed-forward neural network with a set of parameters $\boldsymbol{\theta_{\pi}}$ is employed, which inputs the current estimated state and outputs the policy distribution over all possible learning actions.  

In the RL setup, the state-value function $V^{\pi}(\boldsymbol{s}(t))$ is defined for estimating the expected return of being a given state,
$$V^{\pi}(\boldsymbol{s}(t)) = E^{\pi}[\sum_{i=t}^{T-1}R(i)|\boldsymbol{s}(i)= \boldsymbol{s}(t)],$$
which is the expected total reward when starting from $\boldsymbol{s}(t)$ and following policy $\pi$.
%
To guide the policy learning in a right direction, $V^{\pi}(\boldsymbol{s}(t))$ can be viewed as the baseline to determine whether a specific action leads to high-rewarding knowledge state. Specifically for the critic, we define the advantage function that measures the efficiency of current learning action $a(t)$ by applying the temporal difference (TD) error \citep{sutton1988learning, sutton1998introduction}:
$$A(\boldsymbol{\hat{s}}(t),a(t))= [R(t) + V^{\pi}(\boldsymbol{\hat{s}}(t+1))]-V^{\pi}(\boldsymbol{\hat{s}}(t)),$$
where an empirically observed curiosity reward $R(t)$ is obtained from \eqref{reward}. By subtracting $V^{\pi}(\boldsymbol{\hat{s}}(t))$, $A(\boldsymbol{\hat{s}}(t),a(t))$ calculates the extra reward from taking $a(t)$ beyond the expected value of that state. If $A(\boldsymbol{\hat{s}}(t),a(t))$ is positive, then the tendency to select $a(t)$ in the future should be encouraged and vice versa.

In order to tackle the state-value all the time in the large state space, we employ another feed-forward neural network to approximate the state-value function, that is $V^{\pi}(\boldsymbol{\hat{s}}(t); \boldsymbol{\theta_{v}})$ with a set of parameters $\boldsymbol{\theta_{v}}$ and the input $\boldsymbol{\hat{s}}(t)$. 

\subsubsection{Update} The learner explores the knowledge space following the sampling of the policy function $\pi(a|\boldsymbol{\hat{s}};\boldsymbol{\theta_{\pi}})$, thus accumulating $\{\boldsymbol{\hat{s}}(t), a(t), R(t), \boldsymbol{\hat{s}}(t+1)\}_{t=0}^{T-1}$ for one trajectory. With such entire learning experience data for one learning trajectory, we jointly update the parameterizations of $V^{\pi}(\boldsymbol{\hat{s}}; \boldsymbol{\theta_{v}})$ and $\pi(a|\boldsymbol{\hat{s}};\boldsymbol{\theta_{\pi}})$, which are compatible with each other and alternately implemented in the training.

According to the TD error, the state-value function is essentially updated by minimizing the least squares temporal difference (LSTD; \cite{bradtke1996linear}),
\begin{align*}
	\boldsymbol{\theta_{v}} & = \arg\min_{\boldsymbol{\theta_{v}}}[A(\boldsymbol{\hat{s}}(t),a(t); \boldsymbol{\theta_{v}})]^2\\
	& =  \arg\min_{\boldsymbol{\theta_{v}}}[R(t) + V^{\pi}(\boldsymbol{\hat{s}}(t+1); \boldsymbol{\theta_{v}})-V^{\pi}(\boldsymbol{\hat{s}}(t); \boldsymbol{\theta_{v}})]^2.
\end{align*}
Then the advantage function at time $t$, $A_t$, is directly applied into the policy network for policy gradient \citep{sutton2000policy}, that is
$$\nabla_{\boldsymbol{\theta}_\pi}J(\boldsymbol{\theta_{\pi}}) = E^{\pi}[\sum_{t=0}^{T-1}\nabla_{\boldsymbol{\theta}_\pi}\log\pi(a(t)|\boldsymbol{\hat{s}}(t);\boldsymbol{\theta}_\pi)A_t],$$
and we have $\boldsymbol{\theta}_\pi  \leftarrow \boldsymbol{\theta}_\pi+\nabla_{\boldsymbol{\theta}_\pi}J(\boldsymbol{\theta_{\pi}})$. The update rule improves the recommendation policy iteratively following the critic feedback from the action it generates. Furthermore, subtracting the state-value function plays a role to reduce the high variance inherent in gradient computations, which avoids the policy distribution skewing to a biased direction \citep{williams1992simple}. 

\begin{figure}[htb]
	\centering	
	\includegraphics[width=.46\linewidth]{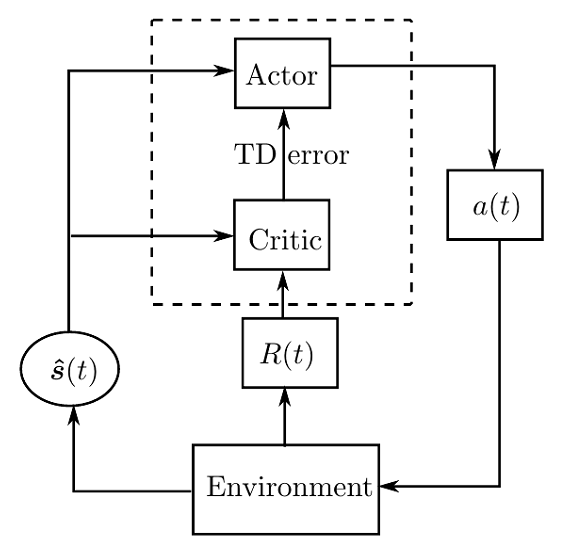}
	\caption{The overview of the policy learning based on the actor-critic framework.  }
	\label{ac}
\end{figure}

The essence of the actor-critic framework is that the actor and critic reinforce each other \citep{williams1992simple}: a correct $\pi(a|\boldsymbol{\hat{s}};\boldsymbol{\theta_{\pi}})$ motivated by curiosity rewards provides high-rewarding learning trajectories to update $V^{\pi}(\boldsymbol{\hat{s}}; \boldsymbol{\theta_{v}})$ towards the right direction; a correct $V^{\pi}(\boldsymbol{\hat{s}}; \boldsymbol{\theta_{v}})$ instructs the correct actions for $\pi(a|\boldsymbol{\hat{s}};\boldsymbol{\theta_{\pi}})$ to reinforce. This is how the agent confirms whether the current learning action is favorable for learning performance improvement. The actor-critic method can be represented schematically as shown in Figure \ref{ac} and we provide the full algorithm including training tricks in Appendix.


\section{4. Experiments}

In this section, we conduct 3 experiments in the increasingly complex and concrete learning scenarios to examine the performance of the proposed curiosity-driven recommendation policy. We not only employ discrete and continuous assessment models but also consider conditional and multi-pointed hierarchical learning paths respectively. Especially, a realistic transition model with the large unlimited knowledge state space is applied and used to simulate learning data in the continuous cases. The results of simulations demonstrate that although the curiosity-driven recommendation strategy pays attention to address the psychological need of learning, it can still achieve a surprising mastery level on knowledge points in the limited time horizon.

\subsection{4.1. Evaluation Criteria} 
Suppose there is a final exam at $T$. It is natural to view the final grade as the evaluation criteria to measure performances of different recommendation strategies.
For the training process, we scale the weighted terminal achievement on the knowledge points at the end of each training episode(i.e., one complete update of the policy given an entire learning trajectory) and denote it as \textit{score}:
$$score =100\times\boldsymbol{w}'\boldsymbol{s}(T),$$ 
where $ \boldsymbol{w} $ is the weight of knowledge points from domain knowledge and $\boldsymbol{s}(T)$ is the terminal knowledge state at $T$. Since we conduct the experiments on synthetic data, the true knowledge state can be obtained at $T$. In the following experiments, we use \textit{score} as the evaluation criteria to evaluate the learning efficiency.

\subsection{4.2. Discrete Case} 
We firstly start from a toy example, which follows a simple one-to-one pointed learning path with discrete binary mastery attributes. Although simple, the simulation setting provides a manifest description about this dynamic decision-making problem.

\subsubsection{Simulation settings} 
\begin{figure}[htb]
	\centering
	\includegraphics[width=.4\linewidth]{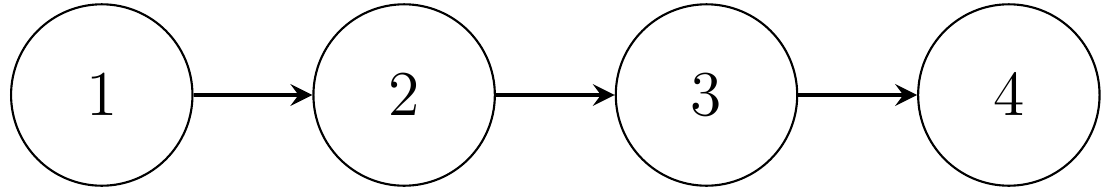}
	\caption{The hierarchy of knowledge points in the Discrete Case, where the number in the circle indicates the corresponding knowledge point and the arrow indicate the prerequisite for learning the pointed knowledge point. }
	\label{hierar2}
\end{figure}

Consider there are four knowledge points $ (K=4) $. Given an initial state $(0,0,0,0)$, the knowledge statue only has binary features $0$ or $1$, corresponding to the non-mastery and mastery of the knowledge point. Specifically, the four knowledge points can be regarded as the following four skills: addition, multiplication, exponentiation and logarithm. One has to master addition before multiplication, and then going for exponentiation and logarithm operation. We summarize such a hierarchical relationship by a one-to-one chain structure in Figure \ref{hierar2}. 
For example, mastering knowledge point 1 is a prerequisite to master knowledge point $ 2 $.

According to the chain structure, there are only five available knowledge states, $(0,0,0,0),(1,0,0,0),(1,1,0,0), (1,1,1,0) , (1,1,1,1) $, denoted as $ \boldsymbol{S}_1, \boldsymbol{S}_2, \boldsymbol{S}_3, \boldsymbol{S}_4 $ and $ \boldsymbol{S}_5 $ respectively. The action space $D$ includes four lecture materials $ D=\{d_1, d_2, d_3, d_4\} $. The transitions matrices given learning actions $ d_1, d_2, d_3$ and $ d_4 $ respectively are shown below:

\begin{equation*}
\boldsymbol{P^{d_1}}={\begin{bmatrix}
	0.5  & 0.5 & 0.0  & 0.0 & 0.0\\
	0.0  & 1.0 & 0.0  & 0.0 & 0.0\\
	0.0  & 0.0 & 1.0  & 0.0 & 0.0\\
	0.0  & 0.0 & 0.0  & 1.0 & 0.0\\
	0.0  & 0.0 & 0.0  & 0.0 & 1.0\\
	\end{bmatrix}},
\boldsymbol{P^{d_2}}={\begin{bmatrix}
	1.0  & 0.0 & 0.0  & 0.0 & 0.0\\
	0.0  & 0.4 & 0.6  & 0.0 & 0.0\\
	0.0  & 0.0 & 1.0  & 0.0 & 0.0\\
	0.0  & 0.0 & 0.0  & 1.0 & 0.0\\
	0.0  & 0.0 & 0.0  & 0.0 & 1.0\\
	\end{bmatrix}},
\end{equation*}
\begin{equation*}
\boldsymbol{P^{d_3}}={\begin{bmatrix}
	1.0  & 0.0 & 0.0  & 0.0 & 0.0\\
	0.0  & 1.0 & 0.0  & 0.0 & 0.0\\
	0.0  & 0.0 & 0.7  & 0.3 & 0.0\\
	0.0  & 0.0 & 0.0  & 1.0 & 0.0\\
	0.0  & 0.0 & 0.0  & 0.0 & 1.0\\
	\end{bmatrix}},
\boldsymbol{P^{d_4}}={\begin{bmatrix}
	1.0  & 0.0 & 0.0  & 0.0 & 0.0\\
	0.0  & 1.0 & 0.0  & 0.0 & 0.0\\
	0.0  & 0.0 & 1.0  & 0.0 & 0.0\\
	0.0  & 0.0 & 0.0  & 0.6 & 0.4\\
	0.0  & 0.0 & 0.0  & 0.0 & 1.0\\
	\end{bmatrix}},
\end{equation*}
where each learning action only relates to one knowledge points. The size of each matrix is $ 5\times 5 $ since there are a total of five possible states. The transition matrices indicate the probabilities of transitions on each knowledge state by taking certain action. For example, the $(2,3)$-entry in $\boldsymbol{P^{d_2}}$ is $ 0.6 $, indicating that the learner at state $ \boldsymbol{S}_2 $ can master knowledge point 2 with probability 0.6 after taking action $ d_2 $. Furthermore, the transition matrices imply the assumption that there is no retrograde during the learning process. Note that the transition model is unknown in the training but used to generate data.
We set $ \boldsymbol{w} = (0.15, 0.25, 0.35, 0.25)' $ and the terminal time $ T = 15 $.

We conduct five simulations in total, including a random policy that takes action uniformly at random from the action space. For other four policies, we apply the proposed curiosity-driven policy and consider the DINA model as the assessment model. Each learning action is followed by an assessment with $ J $ items. The choice of $ J $ is $4, 8$ and $16 $, denoted as DINA\_4, DINA\_8 and DINA\_16 respectively. For the DINA model, the slipping and guessing parameters are generated from a uniform distribution over the interval $[0.1, 0.3]$. A point estimate $\boldsymbol{\hat{s}}$ based on the item responses at each time step is obtained by maximum likelihood estimation.
To study the effect of the assessment error, we also test the proposed policy given the true knowledge state in the simulation, denoted as NO\_DINA. We generate 100 independent replications of each simulation setting and then take the average.
More training details are provided in Appendix.


\subsubsection{Simulation results}

\begin{figure}[htb]
	\centering
	\includegraphics[width=.6\linewidth]{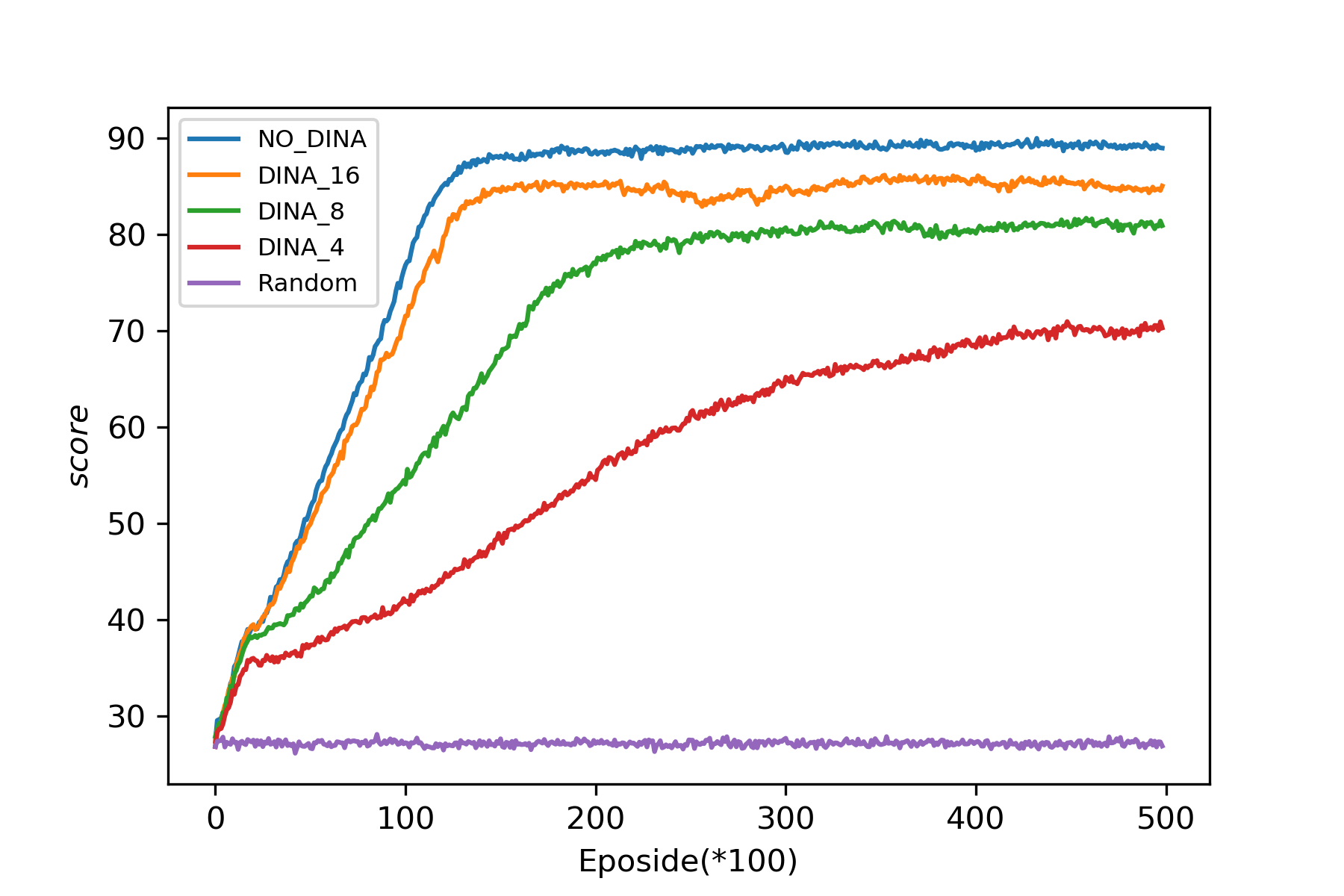}
	\caption{The average \textit{scores} in the Discrete Case based on 100 independent replications of each simulation setting.}
	\label{dis_res}
\end{figure}

We show the average \textit{scores} at the terminal time $T$ along with increasing training episodes in Figure \ref{dis_res}. For ease of display, each point of the curves plots the average \textit{score} for every 100 episodes. 

Clearly, the NO\_DINA curve outperforms others and finally tends to converge with the highest \textit{score}. As to the overall tendency, we can observe increasing \textit{scores} for all settings guided by the proposed policy as training episodes increase. In comparisons of simulations with assessment errors, DINA\_16 that enjoys the most accurate assessments on knowledge points has the best performance that DINA\_4 and DINA\_8 as expected.

\subsection{4.3. Continuous Case I}
Discrete mastery attributes of the knowledge points characterize the knowledge profile in a limited way. \cite{tan2019recommendation} considered a complex environment allowing for the large continuous knowledge state space to imitate a real learning scenario, which is well-designed to model the dynamics of the transition model. In this simulation, we apply the curiosity-driven recommendation strategy in such a challenging environment and demonstrate its performance under the knowledge graph with conditional hierarchical constraints.

\subsubsection{Simulation settings} 
Before introducing the specific setting in the simulation, we first go through the transition model, which is always unknown in the training phase and only used to generate the learning trajectory. Consider learning materials in the action space, some materials training multiple knowledge points simultaneously while some only training single one. For each learning material $a$, define $\boldsymbol{W}_{a}$ as the $K$-dimensional training weight for knowledge points. Given $\boldsymbol{W}_{a(t)}$, the transition with current state $\boldsymbol{s}(t)$ and action $a(t)$ takes the form:
\begin{equation}\label{trans}
\boldsymbol{s}(t+1) = \boldsymbol{1} - (\boldsymbol{1}-\boldsymbol{s}(t))\odot\exp\{-\xi \cdot \boldsymbol{W}_{a(t)} \odot P(\boldsymbol{s}(t)) \},
\end{equation}
where $ \odot $ stands for element-wise multiplication of vectors, $\xi \sim \chi^2_2$, and $P(\boldsymbol{s}(t))$ summarizes the learning prerequisite in terms of the hierarchy in the learning path. Specifically, $P(\cdot)$ maps the knowledge state to a $K$-dimensional zero-one vector where one means the learner is qualified to learn the corresponding knowledge point, zero otherwise. We provide some explanations and features about this transition model below:
\begin{itemize}
\item The equation in \ref{trans} actually describes the learning in such a way that the acquisition of knowledge may be easy during the initial attempts and gradually levels out with less new knowledge gained over time \citep{yelle1979learning}. It means, the transition of knowledge states will be relatively harder as the mastery level gets higher.
\item The randomness of the environment is reflected in $\xi \sim \chi^2_2 $, the degree of freedom of which can be adjustable and used to classify different transition models for different types of learners. For instance, \cite{tan2019recommendation} takes $\xi$ follows $ \chi^2_1 $ and $\chi^2_8 $ to generate data for passive and positive learners respectively. 		
\item For a unlimited continuous knowledge space, it is too strict to determine the transition by simply ascertaining non-mastery or mastery. Beyond the simple hierarchy structure in the discrete case, we impose conditional constraints on mastery attributes of knowledge points. Such constraints come from domain knowledge at this stage, which can be determined by experts according to specific requirements for learners. 
\end{itemize}

\begin{figure}[htb]
	\centering
	\includegraphics[width=.6\linewidth]{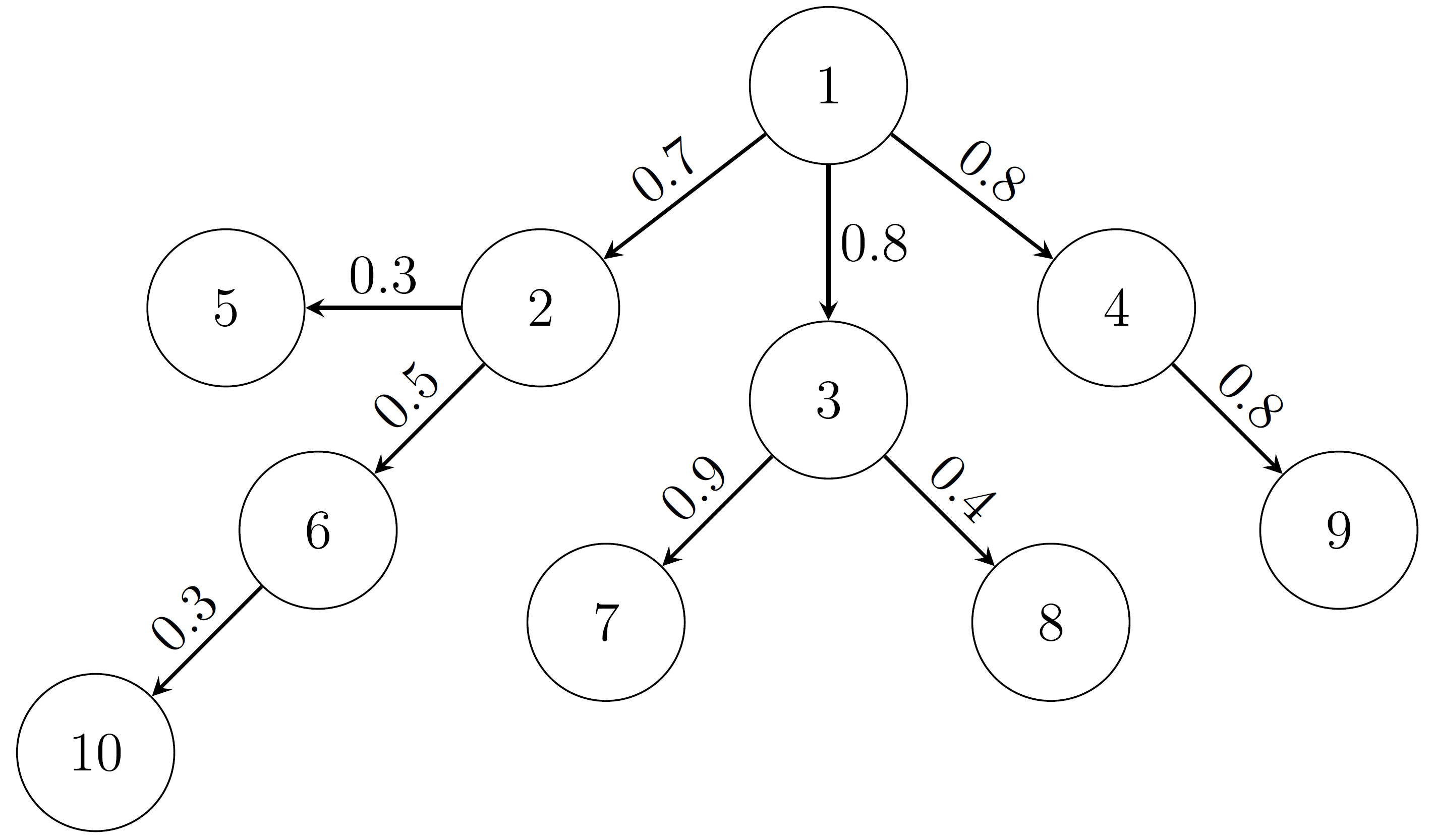}
	\caption{The hierarchy of knowledge points in the Continuous Case I, where the number on each arrow indicates the prerequisite of the mastery attributes before learning certain pointed knowledge points. For example, the mastery attribute of knowledge point 1 has to be no less than 0.7 before learning point 2.}
	\label{con_1}
\end{figure}

\begin{table}
	\scriptsize{ \caption{The corresponding descriptions in the Khan Academy for knowledge points in the Continuous Case I.}
		\label{know1}
		\begin{center}
			\scriptsize
			\begin{tabular}{ccccccccccccccccccccc} \hline
				Knowledge points  &Descriptions\\
				\hline
				$1$ & Dataset warm-up\\
				$2$ & Creating dot plots\\
				$3$ & Calculating the mean\\
				$4$ & Calculating the median\\
				$5$ & Reading dot plots\\
				$6$ & Creating histograms\\
				$7$ & Missing value given mean\\
				$8$ & Effects of shifting, adding, and removing data point\\
				$9$ & Interquantile range(IQR)\\
				$10$ & Reading histograms\\
				 \hline
			\end{tabular}
		\end{center}
	}
\end{table}

\begin{table}
	\scriptsize{ \caption{Learning materials for training certain knowledge points in the Continuous Case I.}
		\label{tab1}
		\begin{center}
			\scriptsize
	\begin{tabular}{cccccccccccccccccccccccccccccccccc} \hline
	Learning materials & $d_1$ & $d_2$ &  $d_3$ & $d_4$ & $d_5$ & $d_6$ & $d_7$ & $d_8$ \\
	Knowledge points to be trained & 1 & 2 &  3 & 4 & 2,5 & 2,6 & 6,10 & 3,7  \\ \hline
	Learning materials & $d_9$ & $d_{10}$ & $d_{11}$ & $d_{12}$ & $d_{13}$ & $d_{14}$ & $d_{15}$ & \\
	Knowledge points to be trained  & 3.8 & 7,8 & 4,9 & 4,9 & 1,2,5 & 1,2,6,10  & 1,3,7,9 & \\ \hline
\end{tabular}
		\end{center}
	}
\end{table}

Specifically, given the initial state $\boldsymbol{s}(0)=(0,0,0,0,0,0,0,0,0,0)'$, the learner have 10 knowledge points($K=10$) to learn within $T=25$. The hierarchical learning path is subject to the constraints shown in Figure \ref{con_1}, which belongs to the knowledge map in the Khan Academy \footnote[1]{https://www.khanacademy.org/exercisedashboard}, an online learning platform. We list detailed descriptions for knowledge points in Table \ref{know1}.  In terms of learning actions, a total of 15 learning materials, i.e., $D=\{d_1,d_2,\dots, d_{15}\}$, are generated and the corresponding knowledge points to be trained by every learning material are presented in Table \ref{tab1}, which are assigned with training weights as $\boldsymbol{W}_{d}$ in the transition model. The weight of knowledge point for evaluation $\boldsymbol{w}=(0.05,0.1,0.05,0.1,0.1,0.2,0.15,0.1,0.1,0.05)'$.
Following every learning action, the M3PL IRT model is adopted to obtain $\boldsymbol{\hat{s}}$ with $ J $ item. The maximum likelihood estimation is used and the mastery attribute of each knowledge point is rescaled to be a continuous value in the interval [0, 1] by the logistic function. We show cases when $ J = 2 $ and $8$ to study different effects of the assessment errors, marked as IRT\_2 and IRT\_8 respectively. Similarly, the proposed policy given the perfect assessment and the random policy are also studied for better comparison, denoted as NO\_IRT and Random. For each simulation setting, we generate 100 independent replications and then take the average to plot.

\subsubsection{Simulation results} 
\begin{figure}[htb]
	\centering
	\includegraphics[width=.6\linewidth]{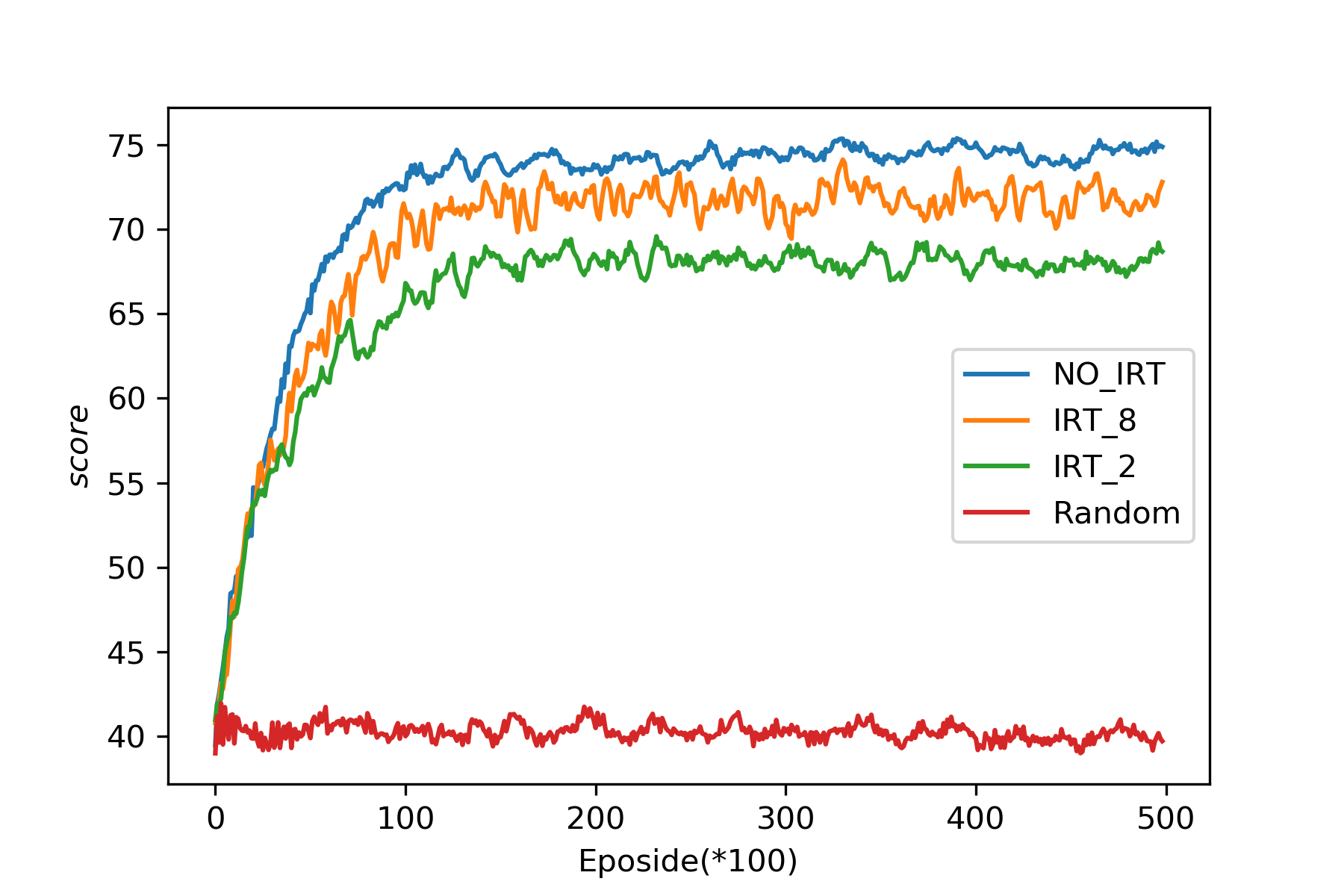}
	\caption{The average \textit{scores} based on 100 independent replications of each simulation setting in the Continuous Case I.}
	\label{con1_res1}
\end{figure}
Figure \ref{con1_res1} shows the result. In each training process we conduct totally 50000 episodes and then draw the average \textit{score} for every 100 episodes.
As we can observe, all experiments following the curiosity-driven policy achieve dominantly higher \textit{scores} than random policy. It substantially demonstrates that the curiosity-driven policy successes in handling a complicated dynamic environment with a large knowledge state space. In particular, the tendencies of IRT\_2, IRT\_8 and NO\_IRT agree with the discrete case, revealing that a more accurate measurement lays a solid foundation for a better recommendation. 


\subsection{4.4. Continuous Case II}
Following the unknown transition model \citep{tan2019recommendation} in the Continuous Case I, we consider a more practical and applicable learning scenario. The hierarchy is designed in such a way that there exist multi-pointed directions in the learning path, which means more than one prerequisites need to be met for learning certain knowledge points.
\subsubsection{Simulation settings}

\begin{figure}[htb]
	\centering
	\includegraphics[width=.6\linewidth]{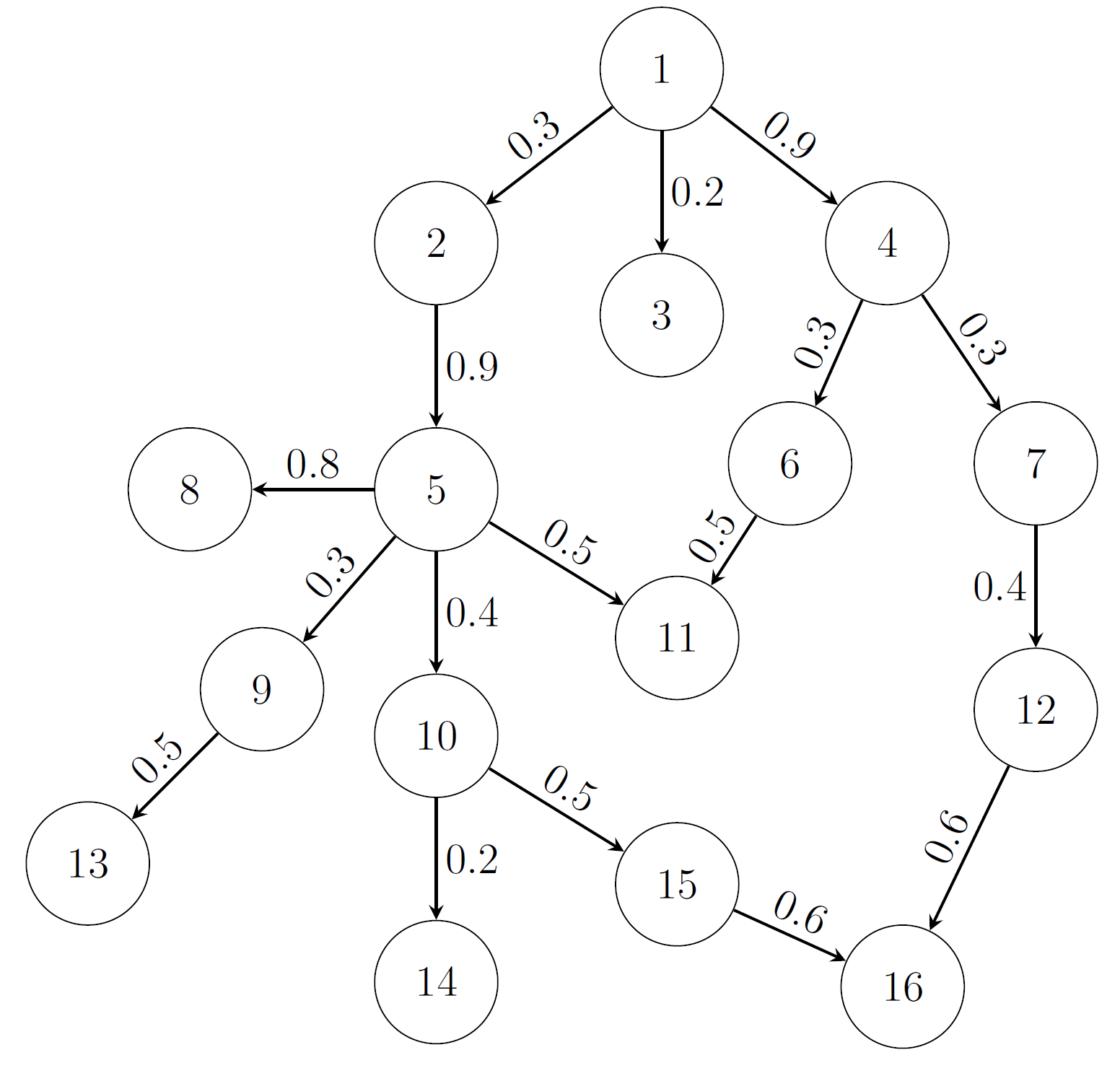}
	\caption{The hierarchy of knowledge points in the Continuous Case II, where the number on each arrow indicates the prerequisite of the mastery attributes before learning certain pointed knowledge points. For example, both the mastery attribute of knowledge point 5 and 6 have to be no less than 0.5 before learning point 11.}
	\label{con_2}
\end{figure}

\begin{table}
	\scriptsize{ \caption{The corresponding descriptions in the Khan Academy for knowledge points in the Continuous Case II.}
		\label{know2}
		\begin{center}
			\scriptsize
			\begin{tabular}{ccccccccccccccccccccccccccccccccc} \hline
				Knowledge points  &Descriptions\\
				\hline
				$1$ & Counting objects 1\\
				$2$ & Making 5\\
				$3$ & Counting object 2\\
				$4$ & Add within 10\\
				$5$ & Add within 5\\
				$6$ & Teen numbers\\
				$7$ & Subtract within 10\\
				$8$ & Making small numbers in different ways\\
				$9$ & Subtract within 5\\
				$10$ & Making 10(grids and number bonds)\\
				$11$ & 2 digital place value challenge\\
				$12$ & Subtraction word problems within 10\\
				$13$ & Relationships between addition and subtraction\\
				$14$ & Making 10\\
				$15$ & Addition word problems within 10\\
				$16$ & Adding\&Subtracting word problems\\
				\hline
			\end{tabular}
		\end{center}
	}
\end{table}
Within time steps $T=40$, the learner have $K=16$ knowledge points to master and the detailed description of knowledge points is shown in Table \ref{know2}. We still adopt the hierarchy among knowledge points from the knowledge map in Khan Academy and add conditional prerequisites as presented in Figure \ref{con_2}. In addition, a total of $22$ learning materials($|D|=22$) are generated to train knowledge points, see Table \ref{tab2}. Instead of assigning a specific knowledge weight, we give every knowledge point with equal weight when calculating \textit{score} for the evaluation.
As with the Continuous Case I, we have the same assessment procedure here and the random policy is also used to compare with all three simulation settings guided by the curiosity-driven recommendation strategy. More training details can be found in Appendix.

\begin{table}
	\scriptsize{ \caption{Learning materials for training certain knowledge points in the Continuous Case II.}
	\label{tab2}
	
	\renewcommand\arraystretch{1}
	\begin{center}
	\scriptsize
	
	\begin{tabular}{cccccccccccccccccccccccccccccccccccccccccccccccccccc} \hline
		Learning materials & $d_1$ & $d_2$ &  $d_3$ & $d_4$ & $d_5$ & $d_6$ & $d_7$ & $d_8$ \\
		Knowledge points to be trained & 1,2 & 1,3 & 1,2,5 & 1,2,3 & 4 & 2,5 & 5,8 & 5,9  \\ \hline
		Learning materials  & $d_9$ & $d_{10}$ & $d_{11}$ & $d_{12}$ & $d_{13}$ & $d_{14}$ & $d_{15}$ & $d_{16}$\\
		Knowledge points to be trained  & 4,6,7 & 5,10 & 10,14 & 5,11 & 9,13 & 10,15 & 15,16 & 7,12,16  \\ \hline
		Learning materials & $d_{17}$ & $d_{18}$ &  $d_{19}$ & $d_{20}$ & $d_{21}$ & $d_{22}$ &  &  \\
		Knowledge points to be trained & 7,12,16 & 10,15,16 & 5,10,14 & 5,9,13 &  5,6,11 & 12,15,16  &  &   \\ \hline
	\end{tabular}
\end{center}
	}
\end{table}

\subsubsection{Simulation results}

\begin{figure}[htb]
	\centering
	\includegraphics[width=.6\linewidth]{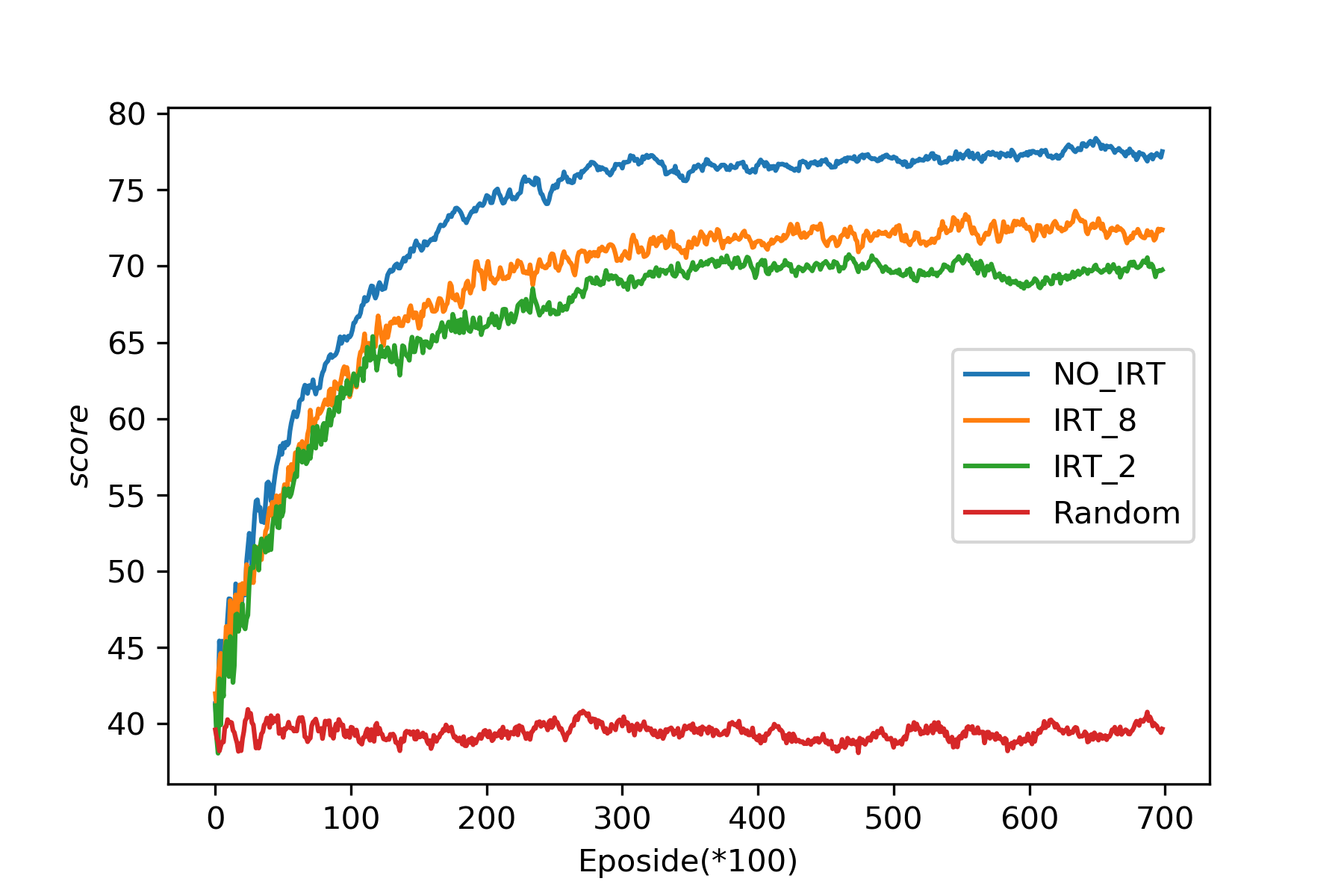}
	\caption{The \textit{scores} based on 100 independent replications of each simulation setting in the Continuous Case II.}
	\label{con2_res1}
\end{figure}


We conducted totally 70000 episodes and provide the result figure here, Figure \ref{con2_res1} showing the average \textit{score} curves based on 100 replications of independent training.
We observe that, the \textit{score} achieved at the terminal time raises up along with the increasing training episodes. Even it seems to be stable, there indeed exist some fluctuations. That is because (1) the randomness with respect to unknown transition model, (2) the stochasticity from outcomes of the policy network and (3) the variability inherent in the gradient optimization. 

Overall, all the \textit{scores} obtained by the proposed policy outperforms the random one, which bears out its  feasibility in the real personalized recommendation problems. 
In practice, with adequate learning data and other complementary information, the proposed method can be reformed to explore the environment better and effectively avoid the interference of environmental noise on the recommendation effects.

%

\section{5. Discussion}
In this paper, we propose a curiosity-driven recommendation strategy for the adaptive learning system. Compared with previous works, the main contribution of this paper is to introduce the curiosity as the engine of recommendation to provide learners with high-rewarding and intrinsically inspired learning instructions. We cast our problem into the reinforcement learning framework. On the one hand, a predictive model predicts the familiarity of knowledge points to partially model the curiosity. On the other hand, we utilize the actor-critic as a model-free method to approximate the optimal policy. The proposed strategy is flexible to scale up and capable to handle a large-scale personalized learning problem. Finally, we showcase discrete and continuous experiments to validate the efficiency of the proposed strategy.

In terms of reward setting, we extract the intrinsic reward entirely from collected learning data rather than extrinsic pre-defined rewards. It requires consecutive learning procedures so as to provide a steady flow of curiosity rewards during training. In addition, considering a complete learning trajectory in our design, we assume that there exists no reward at the terminal time. Actually, it can be more practical if a terminal reward closely related to the mastery levels of knowledge points at the terminal time is obtained, like the final grade. Combining the curiosity rewards with such an extrinsic terminal reward, the policy may be able to receive a more straightforward signal that mastering knowledge points to a good level will lead to higher rewards. In this way, the policy may tend to deliver high-rewarding actions towards an improvement on knowledge states. Domain knowledge can help us build a more purposeful recommendation strategy that focuses on knowledge points with greater weights.

Come back to the policy, we build a predictive model through a simple neural network with current knowledge state and the learning action as inputs.
However, it is hard to directly predict the next state well when the action space is extremely large. The agent is easily attracted to stochastic elements in the environment. Motivated by \cite{burda2018large}, it is better sometimes to transform the raw input into a feature space with only relevant information to avoid noises in the prediction. That is to say, we can extract features from knowledge states and make prediction on those features rather than the entire knowledge statue to better model distinct personalities of learners.


\bibliography{ref} 

\begin{thebibliography}{}

\bibitem [\protect \citeauthoryear {%
Barto%
, Sutton%
\BCBL {}\ \BBA {} Anderson%
}{%
Barto%
\ \protect \BOthers {.}}{%
{\protect \APACyear {1983}}%
}]{%
barto1983neuronlike}
\APACinsertmetastar {%
barto1983neuronlike}%
\begin{APACrefauthors}%
Barto, A\BPBI G.%
, Sutton, R\BPBI S.%
\BCBL {}\ \BBA {} Anderson, C\BPBI W.%
\end{APACrefauthors}%
\unskip\
\newblock
\APACrefYearMonthDay{1983}{}{}.
\newblock
{\BBOQ}\APACrefatitle {Neuronlike adaptive elements that can solve difficult
  learning control problems} {Neuronlike adaptive elements that can solve
  difficult learning control problems}.{\BBCQ}
\newblock
\APACjournalVolNumPages{IEEE transactions on systems, man, and
  cybernetics}{}{5}{834--846}.
\PrintBackRefs{\CurrentBib}

\bibitem [\protect \citeauthoryear {%
Bradtke%
\ \BBA {} Barto%
}{%
Bradtke%
\ \BBA {} Barto%
}{%
{\protect \APACyear {1996}}%
}]{%
bradtke1996linear}
\APACinsertmetastar {%
bradtke1996linear}%
\begin{APACrefauthors}%
Bradtke, S\BPBI J.%
\BCBT {}\ \BBA {} Barto, A\BPBI G.%
\end{APACrefauthors}%
\unskip\
\newblock
\APACrefYearMonthDay{1996}{}{}.
\newblock
{\BBOQ}\APACrefatitle {Linear least-squares algorithms for temporal difference
  learning} {Linear least-squares algorithms for temporal difference
  learning}.{\BBCQ}
\newblock
\APACjournalVolNumPages{Machine learning}{22}{1-3}{33--57}.
\PrintBackRefs{\CurrentBib}

\bibitem [\protect \citeauthoryear {%
Burda%
\ \protect \BOthers {.}}{%
Burda%
\ \protect \BOthers {.}}{%
{\protect \APACyear {2018}}%
}]{%
burda2018large}
\APACinsertmetastar {%
burda2018large}%
\begin{APACrefauthors}%
Burda, Y.%
, Edwards, H.%
, Pathak, D.%
, Storkey, A.%
, Darrell, T.%
\BCBL {}\ \BBA {} Efros, A\BPBI A.%
\end{APACrefauthors}%
\unskip\
\newblock
\APACrefYearMonthDay{2018}{}{}.
\newblock
{\BBOQ}\APACrefatitle {Large-scale study of curiosity-driven learning}
  {Large-scale study of curiosity-driven learning}.{\BBCQ}
\newblock
\APACjournalVolNumPages{arXiv preprint arXiv:1808.04355}{}{}{}.
\PrintBackRefs{\CurrentBib}

\bibitem [\protect \citeauthoryear {%
Chen%
, Li%
, Liu%
\BCBL {}\ \BBA {} Ying%
}{%
Chen%
\ \protect \BOthers {.}}{%
{\protect \APACyear {2018}}%
}]{%
chen2018recommendation}
\APACinsertmetastar {%
chen2018recommendation}%
\begin{APACrefauthors}%
Chen, Y.%
, Li, X.%
, Liu, J.%
\BCBL {}\ \BBA {} Ying, Z.%
\end{APACrefauthors}%
\unskip\
\newblock
\APACrefYearMonthDay{2018}{}{}.
\newblock
{\BBOQ}\APACrefatitle {Recommendation System for Adaptive Learning}
  {Recommendation system for adaptive learning}.{\BBCQ}
\newblock
\APACjournalVolNumPages{Applied Psychological Measurement}{42}{1}{24--41}.
\newblock
\begin{APACrefDOI} \doi{10.1177/0146621617697959} \end{APACrefDOI}
\PrintBackRefs{\CurrentBib}

\bibitem [\protect \citeauthoryear {%
Chentanez%
, Barto%
\BCBL {}\ \BBA {} Singh%
}{%
Chentanez%
\ \protect \BOthers {.}}{%
{\protect \APACyear {2005}}%
}]{%
chentanez2005intrinsically}
\APACinsertmetastar {%
chentanez2005intrinsically}%
\begin{APACrefauthors}%
Chentanez, N.%
, Barto, A\BPBI G.%
\BCBL {}\ \BBA {} Singh, S\BPBI P.%
\end{APACrefauthors}%
\unskip\
\newblock
\APACrefYearMonthDay{2005}{}{}.
\newblock
{\BBOQ}\APACrefatitle {Intrinsically motivated reinforcement learning}
  {Intrinsically motivated reinforcement learning}.{\BBCQ}
\newblock
\BIn{} \APACrefbtitle {Advances in neural information processing systems}
  {Advances in neural information processing systems}\ (\BPGS\ 1281--1288).
\PrintBackRefs{\CurrentBib}

\bibitem [\protect \citeauthoryear {%
Cybenko%
}{%
Cybenko%
}{%
{\protect \APACyear {1989}}%
}]{%
cybenko1989approximation}
\APACinsertmetastar {%
cybenko1989approximation}%
\begin{APACrefauthors}%
Cybenko, G.%
\end{APACrefauthors}%
\unskip\
\newblock
\APACrefYearMonthDay{1989}{}{}.
\newblock
{\BBOQ}\APACrefatitle {Approximation by superpositions of a sigmoidal function}
  {Approximation by superpositions of a sigmoidal function}.{\BBCQ}
\newblock
\APACjournalVolNumPages{Mathematics of control, signals and
  systems}{2}{4}{303--314}.
\PrintBackRefs{\CurrentBib}

\bibitem [\protect \citeauthoryear {%
Goodfellow%
, Bengio%
, Courville%
\BCBL {}\ \BBA {} Bengio%
}{%
Goodfellow%
\ \protect \BOthers {.}}{%
{\protect \APACyear {2016}}%
}]{%
goodfellow2016deep}
\APACinsertmetastar {%
goodfellow2016deep}%
\begin{APACrefauthors}%
Goodfellow, I.%
, Bengio, Y.%
, Courville, A.%
\BCBL {}\ \BBA {} Bengio, Y.%
\end{APACrefauthors}%
\unskip\
\newblock
\APACrefYear{2016}.
\newblock
\APACrefbtitle {Deep learning} {Deep learning}\ (\BVOL~1).
\newblock
\APACaddressPublisher{Cambridge, MA}{MIT press Cambridge}.
\PrintBackRefs{\CurrentBib}

\bibitem [\protect \citeauthoryear {%
Junker%
\ \BBA {} Sijtsma%
}{%
Junker%
\ \BBA {} Sijtsma%
}{%
{\protect \APACyear {2001}}%
}]{%
junker2001cognitive}
\APACinsertmetastar {%
junker2001cognitive}%
\begin{APACrefauthors}%
Junker, B\BPBI W.%
\BCBT {}\ \BBA {} Sijtsma, K.%
\end{APACrefauthors}%
\unskip\
\newblock
\APACrefYearMonthDay{2001}{}{}.
\newblock
{\BBOQ}\APACrefatitle {Cognitive assessment models with few assumptions, and
  connections with nonparametric item response theory} {Cognitive assessment
  models with few assumptions, and connections with nonparametric item response
  theory}.{\BBCQ}
\newblock
\APACjournalVolNumPages{Applied Psychological Measurement}{25}{3}{258--272}.
\PrintBackRefs{\CurrentBib}

\bibitem [\protect \citeauthoryear {%
Kaelbling%
, Littman%
\BCBL {}\ \BBA {} Moore%
}{%
Kaelbling%
\ \protect \BOthers {.}}{%
{\protect \APACyear {1996}}%
}]{%
kaelbling1996reinforcement}
\APACinsertmetastar {%
kaelbling1996reinforcement}%
\begin{APACrefauthors}%
Kaelbling, L\BPBI P.%
, Littman, M\BPBI L.%
\BCBL {}\ \BBA {} Moore, A\BPBI W.%
\end{APACrefauthors}%
\unskip\
\newblock
\APACrefYearMonthDay{1996}{}{}.
\newblock
{\BBOQ}\APACrefatitle {Reinforcement learning: A survey} {Reinforcement
  learning: A survey}.{\BBCQ}
\newblock
\APACjournalVolNumPages{Journal of Artificial Intelligence
  Research}{4}{}{237--285}.
\newblock
\begin{APACrefDOI} \doi{10.1613/jair.301} \end{APACrefDOI}
\PrintBackRefs{\CurrentBib}

\bibitem [\protect \citeauthoryear {%
Kingma%
\ \BBA {} Ba%
}{%
Kingma%
\ \BBA {} Ba%
}{%
{\protect \APACyear {2014}}%
}]{%
kingma2014adam}
\APACinsertmetastar {%
kingma2014adam}%
\begin{APACrefauthors}%
Kingma, D\BPBI P.%
\BCBT {}\ \BBA {} Ba, J.%
\end{APACrefauthors}%
\unskip\
\newblock
\APACrefYearMonthDay{2014}{}{}.
\newblock
{\BBOQ}\APACrefatitle {Adam: A method for stochastic optimization} {Adam: A
  method for stochastic optimization}.{\BBCQ}
\newblock
\APACjournalVolNumPages{ArXiv preprint arXiv:1412.6980}{}{}{}.
\newblock
\begin{APACrefURL} \url{http://arxiv.org/abs/1412.6980} \end{APACrefURL}
\PrintBackRefs{\CurrentBib}

\bibitem [\protect \citeauthoryear {%
Konda%
\ \BBA {} Tsitsiklis%
}{%
Konda%
\ \BBA {} Tsitsiklis%
}{%
{\protect \APACyear {2000}}%
}]{%
konda2000actor}
\APACinsertmetastar {%
konda2000actor}%
\begin{APACrefauthors}%
Konda, V\BPBI R.%
\BCBT {}\ \BBA {} Tsitsiklis, J\BPBI N.%
\end{APACrefauthors}%
\unskip\
\newblock
\APACrefYearMonthDay{2000}{}{}.
\newblock
{\BBOQ}\APACrefatitle {Actor-critic algorithms} {Actor-critic
  algorithms}.{\BBCQ}
\newblock
\BIn{} \APACrefbtitle {Advances in neural information processing systems}
  {Advances in neural information processing systems}\ (\BPGS\ 1008--1014).
\PrintBackRefs{\CurrentBib}

\bibitem [\protect \citeauthoryear {%
Li%
, Xu%
, Zhang%
\BCBL {}\ \BBA {} Chang%
}{%
Li%
\ \protect \BOthers {.}}{%
{\protect \APACyear {2018}}%
}]{%
li2018optimal}
\APACinsertmetastar {%
li2018optimal}%
\begin{APACrefauthors}%
Li, X.%
, Xu, H.%
, Zhang, J.%
\BCBL {}\ \BBA {} Chang, H\BHBI h.%
\end{APACrefauthors}%
\unskip\
\newblock
\APACrefYearMonthDay{2018}{}{}.
\newblock
{\BBOQ}\APACrefatitle {Optimal Hierarchical Learning Path Design with
  Reinforcement Learning} {Optimal hierarchical learning path design with
  reinforcement learning}.{\BBCQ}
\newblock
\APACjournalVolNumPages{arXiv preprint arXiv:1810.05347}{}{}{}.
\PrintBackRefs{\CurrentBib}

\bibitem [\protect \citeauthoryear {%
Loewenstein%
}{%
Loewenstein%
}{%
{\protect \APACyear {1994}}%
}]{%
loewenstein1994psychology}
\APACinsertmetastar {%
loewenstein1994psychology}%
\begin{APACrefauthors}%
Loewenstein, G.%
\end{APACrefauthors}%
\unskip\
\newblock
\APACrefYearMonthDay{1994}{}{}.
\newblock
{\BBOQ}\APACrefatitle {The psychology of curiosity: A review and
  reinterpretation.} {The psychology of curiosity: A review and
  reinterpretation.}{\BBCQ}
\newblock
\APACjournalVolNumPages{Psychological bulletin}{116}{1}{75}.
\PrintBackRefs{\CurrentBib}

\bibitem [\protect \citeauthoryear {%
Mnih%
\ \protect \BOthers {.}}{%
Mnih%
\ \protect \BOthers {.}}{%
{\protect \APACyear {2016}}%
}]{%
mnih2016asynchronous}
\APACinsertmetastar {%
mnih2016asynchronous}%
\begin{APACrefauthors}%
Mnih, V.%
, Badia, A\BPBI P.%
, Mirza, M.%
, Graves, A.%
, Lillicrap, T.%
, Harley, T.%
\BDBL {}Kavukcuoglu, K.%
\end{APACrefauthors}%
\unskip\
\newblock
\APACrefYearMonthDay{2016}{}{}.
\newblock
{\BBOQ}\APACrefatitle {Asynchronous methods for deep reinforcement learning}
  {Asynchronous methods for deep reinforcement learning}.{\BBCQ}
\newblock
\BIn{} \APACrefbtitle {International conference on machine learning}
  {International conference on machine learning}\ (\BPGS\ 1928--1937).
\PrintBackRefs{\CurrentBib}

\bibitem [\protect \citeauthoryear {%
Mnih%
\ \protect \BOthers {.}}{%
Mnih%
\ \protect \BOthers {.}}{%
{\protect \APACyear {2013}}%
}]{%
mnih2013playing}
\APACinsertmetastar {%
mnih2013playing}%
\begin{APACrefauthors}%
Mnih, V.%
, Kavukcuoglu, K.%
, Silver, D.%
, Graves, A.%
, Antonoglou, I.%
, Wierstra, D.%
\BCBL {}\ \BBA {} Riedmiller, M.%
\end{APACrefauthors}%
\unskip\
\newblock
\APACrefYearMonthDay{2013}{}{}.
\newblock
{\BBOQ}\APACrefatitle {Playing atari with deep reinforcement learning} {Playing
  atari with deep reinforcement learning}.{\BBCQ}
\newblock
\APACjournalVolNumPages{ArXiv preprint arXiv:1312.5602}{}{}{}.
\newblock
\begin{APACrefURL} \url{http://arxiv.org/abs/1312.5602} \end{APACrefURL}
\PrintBackRefs{\CurrentBib}

\bibitem [\protect \citeauthoryear {%
Mnih%
\ \protect \BOthers {.}}{%
Mnih%
\ \protect \BOthers {.}}{%
{\protect \APACyear {2015}}%
}]{%
mnih2015human}
\APACinsertmetastar {%
mnih2015human}%
\begin{APACrefauthors}%
Mnih, V.%
, Kavukcuoglu, K.%
, Silver, D.%
, Rusu, A\BPBI A.%
, Veness, J.%
, Bellemare, M\BPBI G.%
\BDBL {}Hassabis, D.%
\end{APACrefauthors}%
\unskip\
\newblock
\APACrefYearMonthDay{2015}{}{}.
\newblock
{\BBOQ}\APACrefatitle {Human-level control through deep reinforcement learning}
  {Human-level control through deep reinforcement learning}.{\BBCQ}
\newblock
\APACjournalVolNumPages{Nature}{518}{7540}{529--533}.
\newblock
\begin{APACrefDOI} \doi{10.1038/nature14236} \end{APACrefDOI}
\PrintBackRefs{\CurrentBib}

\bibitem [\protect \citeauthoryear {%
Pathak%
, Agrawal%
, Efros%
\BCBL {}\ \BBA {} Darrell%
}{%
Pathak%
\ \protect \BOthers {.}}{%
{\protect \APACyear {2017}}%
}]{%
pathak2017curiosity}
\APACinsertmetastar {%
pathak2017curiosity}%
\begin{APACrefauthors}%
Pathak, D.%
, Agrawal, P.%
, Efros, A\BPBI A.%
\BCBL {}\ \BBA {} Darrell, T.%
\end{APACrefauthors}%
\unskip\
\newblock
\APACrefYearMonthDay{2017}{}{}.
\newblock
{\BBOQ}\APACrefatitle {Curiosity-driven exploration by self-supervised
  prediction} {Curiosity-driven exploration by self-supervised
  prediction}.{\BBCQ}
\newblock
\BIn{} \APACrefbtitle {Proceedings of the IEEE Conference on Computer Vision
  and Pattern Recognition Workshops} {Proceedings of the ieee conference on
  computer vision and pattern recognition workshops}\ (\BPGS\ 16--17).
\PrintBackRefs{\CurrentBib}

\bibitem [\protect \citeauthoryear {%
Powell%
}{%
Powell%
}{%
{\protect \APACyear {2007}}%
}]{%
powell2007approximate}
\APACinsertmetastar {%
powell2007approximate}%
\begin{APACrefauthors}%
Powell, W\BPBI B.%
\end{APACrefauthors}%
\unskip\
\newblock
\APACrefYear{2007}.
\newblock
\APACrefbtitle {Approximate Dynamic Programming: Solving the Curses of
  Dimensionality (Wiley Series in Probability and Statistics)} {Approximate
  dynamic programming: Solving the curses of dimensionality (wiley series in
  probability and statistics)}.
\newblock
\APACaddressPublisher{New York, NY}{Wiley-Interscience}.
\newblock
\begin{APACrefDOI} \doi{10.1002/9780470182963} \end{APACrefDOI}
\PrintBackRefs{\CurrentBib}

\bibitem [\protect \citeauthoryear {%
Reckase%
}{%
Reckase%
}{%
{\protect \APACyear {2009}}%
}]{%
reckase2009multidimensional}
\APACinsertmetastar {%
reckase2009multidimensional}%
\begin{APACrefauthors}%
Reckase, M\BPBI D.%
\end{APACrefauthors}%
\unskip\
\newblock
\APACrefYear{2009}.
\newblock
\APACrefbtitle {Multidimensional item response theory} {Multidimensional item
  response theory}\ (\BVOL~150).
\newblock
\APACaddressPublisher{New York, NY}{Springer}.
\PrintBackRefs{\CurrentBib}

\bibitem [\protect \citeauthoryear {%
Schmidhuber%
}{%
Schmidhuber%
}{%
{\protect \APACyear {1991}}%
}]{%
schmidhuber1991curious}
\APACinsertmetastar {%
schmidhuber1991curious}%
\begin{APACrefauthors}%
Schmidhuber, J.%
\end{APACrefauthors}%
\unskip\
\newblock
\APACrefYearMonthDay{1991}{}{}.
\newblock
{\BBOQ}\APACrefatitle {Curious model-building control systems} {Curious
  model-building control systems}.{\BBCQ}
\newblock
\BIn{} \APACrefbtitle {[Proceedings] 1991 IEEE International Joint Conference
  on Neural Networks} {[proceedings] 1991 ieee international joint conference
  on neural networks}\ (\BPGS\ 1458--1463).
\PrintBackRefs{\CurrentBib}

\bibitem [\protect \citeauthoryear {%
Silver%
\ \protect \BOthers {.}}{%
Silver%
\ \protect \BOthers {.}}{%
{\protect \APACyear {2016}}%
}]{%
silver2016mastering}
\APACinsertmetastar {%
silver2016mastering}%
\begin{APACrefauthors}%
Silver, D.%
, Huang, A.%
, Maddison, C.%
, Guez, A.%
, Sifre, L.%
, van~den Driessche, G.%
\BDBL {}Hassabis, D.%
\end{APACrefauthors}%
\unskip\
\newblock
\APACrefYearMonthDay{2016}{}{}.
\newblock
{\BBOQ}\APACrefatitle {Mastering the game of Go with deep neural networks and
  tree search} {Mastering the game of go with deep neural networks and tree
  search}.{\BBCQ}
\newblock
\APACjournalVolNumPages{Nature}{529}{7587}{484--489}.
\newblock
\begin{APACrefDOI} \doi{10.1038/nature16961} \end{APACrefDOI}
\PrintBackRefs{\CurrentBib}

\bibitem [\protect \citeauthoryear {%
Sleeman%
\ \BBA {} Brown%
}{%
Sleeman%
\ \BBA {} Brown%
}{%
{\protect \APACyear {1982}}%
}]{%
sleeman1982intelligent}
\APACinsertmetastar {%
sleeman1982intelligent}%
\begin{APACrefauthors}%
Sleeman, D.%
\BCBT {}\ \BBA {} Brown, J\BPBI S.%
\end{APACrefauthors}%
\unskip\
\newblock
\APACrefYear{1982}.
\newblock
\APACrefbtitle {{Intelligent Tutoring Systems}} {{Intelligent Tutoring
  Systems}}.
\newblock
\APACaddressPublisher{London, England}{Academic Press}.
\newblock
\begin{APACrefURL} \url{https://hal.archives-ouvertes.fr/hal-00702997}
  \end{APACrefURL}
\PrintBackRefs{\CurrentBib}

\bibitem [\protect \citeauthoryear {%
Sutton%
}{%
Sutton%
}{%
{\protect \APACyear {1988}}%
}]{%
sutton1988learning}
\APACinsertmetastar {%
sutton1988learning}%
\begin{APACrefauthors}%
Sutton, R\BPBI S.%
\end{APACrefauthors}%
\unskip\
\newblock
\APACrefYearMonthDay{1988}{}{}.
\newblock
{\BBOQ}\APACrefatitle {Learning to predict by the methods of temporal
  differences} {Learning to predict by the methods of temporal
  differences}.{\BBCQ}
\newblock
\APACjournalVolNumPages{Machine learning}{3}{1}{9--44}.
\PrintBackRefs{\CurrentBib}

\bibitem [\protect \citeauthoryear {%
Sutton%
, Barto%
\BCBL {}\ \protect \BOthers {.}}{%
Sutton%
\ \protect \BOthers {.}}{%
{\protect \APACyear {1998}}%
}]{%
sutton1998introduction}
\APACinsertmetastar {%
sutton1998introduction}%
\begin{APACrefauthors}%
Sutton, R\BPBI S.%
, Barto, A\BPBI G.%
\BCBL {}\ \BOthersPeriod {.}\end{APACrefauthors}%
\unskip\
\newblock
\APACrefYear{1998}.
\newblock
\APACrefbtitle {Introduction to reinforcement learning} {Introduction to
  reinforcement learning}\ (\BVOL~135).
\newblock
\APACaddressPublisher{}{MIT press Cambridge}.
\PrintBackRefs{\CurrentBib}

\bibitem [\protect \citeauthoryear {%
Sutton%
, McAllester%
, Singh%
\BCBL {}\ \BBA {} Mansour%
}{%
Sutton%
\ \protect \BOthers {.}}{%
{\protect \APACyear {1999}}%
}]{%
sutton2000policy}
\APACinsertmetastar {%
sutton2000policy}%
\begin{APACrefauthors}%
Sutton, R\BPBI S.%
, McAllester, D.%
, Singh, S.%
\BCBL {}\ \BBA {} Mansour, Y.%
\end{APACrefauthors}%
\unskip\
\newblock
\APACrefYearMonthDay{1999}{}{}.
\newblock
{\BBOQ}\APACrefatitle {Policy Gradient Methods for Reinforcement Learning with
  Function Approximation} {Policy gradient methods for reinforcement learning
  with function approximation}.{\BBCQ}
\newblock
\BIn{} \APACrefbtitle {Proceedings of the 12th International Conference on
  Neural Information Processing Systems} {Proceedings of the 12th international
  conference on neural information processing systems}\ (\BPGS\ 1057--1063).
\newblock
\APACaddressPublisher{Cambridge, MA}{MIT Press}.
\newblock
\begin{APACrefURL} \url{http://dl.acm.org/citation.cfm?id=3009657.3009806}
  \end{APACrefURL}
\PrintBackRefs{\CurrentBib}

\bibitem [\protect \citeauthoryear {%
Tan%
, Han%
, Ye%
\BCBL {}\ \BBA {} Chen%
}{%
Tan%
\ \protect \BOthers {.}}{%
{\protect \APACyear {2019}}%
}]{%
tan2019recommendation}
\APACinsertmetastar {%
tan2019recommendation}%
\begin{APACrefauthors}%
Tan, C.%
, Han, R.%
, Ye, R.%
\BCBL {}\ \BBA {} Chen, K.%
\end{APACrefauthors}%
\unskip\
\newblock
\APACrefYearMonthDay{2019}{}{}.
\newblock
{\BBOQ}\APACrefatitle {Adaptive Learning Recommendation Strategy Based on Deep
  Q-learning} {Adaptive learning recommendation strategy based on deep
  q-learning}.{\BBCQ}
\newblock
\APACjournalVolNumPages{Applied Psychological Measurement}{}{}{}.
\newblock
\begin{APACrefDOI} \doi{10.1177/0146621619858674} \end{APACrefDOI}
\PrintBackRefs{\CurrentBib}

\bibitem [\protect \citeauthoryear {%
Tang%
, Chen%
, Li%
, Liu%
\BCBL {}\ \BBA {} Ying%
}{%
Tang%
\ \protect \BOthers {.}}{%
{\protect \APACyear {2019}}%
}]{%
doi:10.1111/bmsp.12144}
\APACinsertmetastar {%
doi:10.1111/bmsp.12144}%
\begin{APACrefauthors}%
Tang, X.%
, Chen, Y.%
, Li, X.%
, Liu, J.%
\BCBL {}\ \BBA {} Ying, Z.%
\end{APACrefauthors}%
\unskip\
\newblock
\APACrefYearMonthDay{2019}{}{}.
\newblock
{\BBOQ}\APACrefatitle {A reinforcement learning approach to personalized
  learning recommendation systems} {A reinforcement learning approach to
  personalized learning recommendation systems}.{\BBCQ}
\newblock
\APACjournalVolNumPages{British Journal of Mathematical and Statistical
  Psychology}{72}{1}{108--135}.
\newblock
\begin{APACrefDOI} \doi{10.1111/bmsp.12144} \end{APACrefDOI}
\PrintBackRefs{\CurrentBib}

\bibitem [\protect \citeauthoryear {%
Thrun%
\ \BBA {} Schwartz%
}{%
Thrun%
\ \BBA {} Schwartz%
}{%
{\protect \APACyear {1993}}%
}]{%
thrun1993issues}
\APACinsertmetastar {%
thrun1993issues}%
\begin{APACrefauthors}%
Thrun, S.%
\BCBT {}\ \BBA {} Schwartz, A.%
\end{APACrefauthors}%
\unskip\
\newblock
\APACrefYearMonthDay{1993}{}{}.
\newblock
{\BBOQ}\APACrefatitle {Issues in Using Function Approximation for Reinforcement
  Learning} {Issues in using function approximation for reinforcement
  learning}.{\BBCQ}
\newblock
\BIn{} D\BPBI T\BPBI J\BPBI E.~M.~Mozer P.~Smolensky\ \BBA {} A.~Weigend\
  (\BEDS), \APACrefbtitle {Proceedings of the 1993 Connectionist Models Summer
  School.} {Proceedings of the 1993 connectionist models summer school.}
\newblock
\APACaddressPublisher{Mahwah, NJ}{Erlbaum Associates}.
\PrintBackRefs{\CurrentBib}

\bibitem [\protect \citeauthoryear {%
Wenger%
}{%
Wenger%
}{%
{\protect \APACyear {1987}}%
}]{%
Wenger1987}
\APACinsertmetastar {%
Wenger1987}%
\begin{APACrefauthors}%
Wenger, E.%
\end{APACrefauthors}%
\unskip\
\newblock
\APACrefYear{1987}.
\newblock
\APACrefbtitle {Artificial Intelligence and Tutoring Systems: Computational and
  Cognitive Approaches to the Communication of Knowledge} {Artificial
  intelligence and tutoring systems: Computational and cognitive approaches to
  the communication of knowledge}.
\newblock
\APACaddressPublisher{San Francisco, CA}{Morgan Kaufmann Publishers Inc.}
\PrintBackRefs{\CurrentBib}

\bibitem [\protect \citeauthoryear {%
Williams%
}{%
Williams%
}{%
{\protect \APACyear {1992}}%
}]{%
williams1992simple}
\APACinsertmetastar {%
williams1992simple}%
\begin{APACrefauthors}%
Williams, R\BPBI J.%
\end{APACrefauthors}%
\unskip\
\newblock
\APACrefYearMonthDay{1992}{}{}.
\newblock
{\BBOQ}\APACrefatitle {Simple statistical gradient-following algorithms for
  connectionist reinforcement learning} {Simple statistical gradient-following
  algorithms for connectionist reinforcement learning}.{\BBCQ}
\newblock
\APACjournalVolNumPages{Machine learning}{8}{3-4}{229--256}.
\PrintBackRefs{\CurrentBib}

\bibitem [\protect \citeauthoryear {%
Yelle%
}{%
Yelle%
}{%
{\protect \APACyear {1979}}%
}]{%
yelle1979learning}
\APACinsertmetastar {%
yelle1979learning}%
\begin{APACrefauthors}%
Yelle, L\BPBI E.%
\end{APACrefauthors}%
\unskip\
\newblock
\APACrefYearMonthDay{1979}{}{}.
\newblock
{\BBOQ}\APACrefatitle {The learning curve: Historical review and comprehensive
  survey} {The learning curve: Historical review and comprehensive
  survey}.{\BBCQ}
\newblock
\APACjournalVolNumPages{Decision sciences}{10}{2}{302--328}.
\PrintBackRefs{\CurrentBib}

\end{thebibliography}

\appendix
\section{}

\begin{algorithm}[htb]
	\caption{ Curiosity-Driven Recommendation Algorithm}
	\centering
	\label{Actor-critic}
	\begin{algorithmic}[5]
		\Require
		Random global network parameters $ \boldsymbol{\theta^g} = (\boldsymbol{\theta_v^g}, \boldsymbol{\theta_\pi^g}, \boldsymbol{\theta_p^g}) $;
		Total episode $ M $;
		\Ensure
		Policy function $\pi$;
		\State Initialize replay memory capacity $ D $ to capacity $N$;
		\For{episode = 1,$ \ldots, M $}
		\State Select one thread which is not working and initialize knowledge state $\boldsymbol{\hat{s}}(0) = \boldsymbol{s}(0)$;
		\State Set the thread-specific parameters $ \boldsymbol{\theta} = \boldsymbol{\theta^g} $, where $ \boldsymbol{\theta} = (\boldsymbol{\theta_v}, \boldsymbol{\theta_\pi}, \boldsymbol{\theta_p})$;
		\For{t = 0, $ \ldots, T-1 $}
		\State Select an action $ a(t) $ according to policy function $ \pi(a(t)|\boldsymbol{\hat{s}}(t);\boldsymbol{\theta_\pi}) $;
		\State Execute action $ a(t) $ in emulator and estimate the next knowledge state $ \boldsymbol{\hat{s}}(t+1) $ through assessment model;
		\State Compute reward $ R(t) = \lVert f(\boldsymbol{\hat{s}}(t), a(t)) - \boldsymbol{\hat{s}}(t+1) \lVert_2^2 $ through the predictive function $ f(\cdot); $
		\State Store transition $ (\boldsymbol{\hat{s}}(t), a(t), \boldsymbol{\hat{s}}(t+1) ) $ in $ D $;
		\State Sample random mini-batch with size $ n $ of transitions $ (\boldsymbol{\hat{s}}(j), a(j), \boldsymbol{\hat{s}}(j+1) ) $ from $ D $;
		\State Set $ d\boldsymbol{\theta_p}  = \sum_{j=1}^{n} \nabla_{\boldsymbol{\theta_p}} \lVert f(\boldsymbol{\hat{s}}(j), a(i); \boldsymbol{\theta_p}) - \boldsymbol{\hat{s}}(j+1)\lVert_2^2; $
		\State Perform asynchronous updates of $\boldsymbol{\theta_p^g} $ and $ \boldsymbol{\theta_p} $ using $ d\boldsymbol{\theta_p} $.
		
		\EndFor \State {\textbf{Until terminal $ \boldsymbol{\hat{s}}(t) $ or $ t = T - 1$}}
		\State Set 	$r(\boldsymbol{\hat{s}}(t)) = \begin{cases} 
		0 & \text{for terminal state } \boldsymbol{\hat{s}}(t)\\
		V(\boldsymbol{\hat{s}}(t);\boldsymbol{\theta_v})& \text{otherwise }
		\end{cases}$
		
		\For{$ i \in \{T-1, \ldots, 0\} $}
		\State Set $r(\boldsymbol{\hat{s}}(i)) = R(i) + r(\boldsymbol{\hat{s}}(i+1))$; 
		\EndFor \State {\textbf{end for}}
		\State Set $ d\boldsymbol{\theta_v} = \sum_{i=0}^{t-1} \nabla_{\boldsymbol{\theta_v}}[(r(\boldsymbol{\hat{s}}(i)) - V(\boldsymbol{\hat{s}}(i);\boldsymbol{\theta_v}))^2];   $ 
		\State Set $ d\boldsymbol{\theta_\pi}  = \sum_{i=0}^{t-1}\nabla_{\boldsymbol{\theta_\pi}} [\log \pi(a(i)|\boldsymbol{\hat{s}}(i);\boldsymbol{\theta_\pi})(r(\boldsymbol{\hat{s}}(i)) - V(\boldsymbol{\hat{s}}(i);\boldsymbol{\theta_v}))]; $
		\State Perform asynchronous update of $ \boldsymbol{\theta_v^g}$ and $\boldsymbol{\theta_\pi^g}$ using $ d\boldsymbol{\theta_v}$ and $d\boldsymbol{\theta_\pi} $ respectively.
		\EndFor \State {\textbf{end for}}

		\Return $ \pi(a(t)|\boldsymbol{s}(t);\boldsymbol{\theta_\pi^g}) $;
	\end{algorithmic}
\end{algorithm}

\subsubsection{Remark} We provide some training details below.  \\
\begin{APAitemize}
	\item The architectures of $f(\cdot)$, the predictive network, are different in the discrete and continuous cases. In the discrete case, the network has two hidden layers while it has three hidden layers in the continuous cases. The rectified linear unit (ReLU), $ReLU(x) = max(0, x)$, serves as the activation function.
	
    \item The architectures of the actor and critic networks are the same, which consists of three hidden layers and the rectified linear unit (ReLU) serves as the activation function. Although simple, the network structure is sufficient to fit the simulated dataset and the architecture can be scaled up for different learning scenarios. 
    
	\item Hyperparameters setting in the training: memory capacity $N= 6000 $, batch size $n=64$.

	\item In the discrete case, the learning rate of the actor and critic networks is $0.0006$ while the learning rate of the prediction model is $ 0.006 $. We conduct $ 50000 $ episodes in the training phase.
		
	\item In the continuous cases, the learning rate of the actor and critic networks is $0.0005$ while the learning rate of the prediction model is $ 0.002 $. We conduct $ 50000 $ and $70000$ episodes for Continuous Case I and II respectively in the training phase.
	
	\item $\boldsymbol{\theta}$ is updated for each iteration by the method of stochastic gradient descent with Adam \citep{kingma2014adam}. 
	
	\item To efficiently explore the state space, A3C \citep{mnih2016asynchronous} is employed to implement training where multiple workers in parallel environments independently update a global value function. 

\end{APAitemize}

\end{document}